\documentstyle{article}

\begin{document}
\title{The Initialization Problem in Quantum Computing}
\author{Subhash Kak\\
Department of Electrical \& Computer Engineering\\
Louisiana State University\\
Baton Rouge, LA 70803-5901; {\tt kak@ece.lsu.edu}}
\date{November 29, 1999}
\maketitle

\begin{abstract}
The problem of initializing phase in a quantum computing system is considered.
The initialization of phases
is a problem when the system is initially present in a superposition
state as well as 
in the application of the
quantum gate transformations, since each gate will
introduce phase uncertainty.
The accumulation of these random phases will
reduce the effectiveness of
the recently proposed quantum computing schemes.
The paper also presents general
observations on the nonlocal nature of quantum errors and
the expected
performance of the
proposed quantum error-correction codes that
are based on the assumption that the errors are
either
bit-flip or phase-flip or both.
It is argued that these codes cannot directly solve the initialization
problem of quantum computing.


\end{abstract}

\vspace{3mm}
\subsection*{\centering Foundations of Physics, vol 29, 1999, pp. 267-279}

\newpage

\section{Introduction}
A quantum state is undetermined with respect to its phase.
This indeterminacy is in principle irremovable\cite{La77}.
The uncertainty of phase, together with superposition, is
responsible for the power of quantum mechanics.
It also compels us to speak of information in
{\it positivist} terms---with
respect to an observation rather than in an absolute sense.

Since phase indeterminacy is fundamental to quantum
description, it is relevant to examine its
implications for quantum computation.
This indeterminacy can manifest itself in a variety
of ways due to the interaction with the environment 
or 
while initializing the quantum register.
Quantum computing algorithms
assume that the state of the quantum register
has its phase uncertainty lumped together, so that it can
be ignored.
This is true enough in certain idealized state preparations.
But more realistic situations may not
permit it to be
lumped together.

Effects of decoherence in
the implementation of quantum
computation have been widely discussed in the
literature\cite{Ek96,Ge97,Pl96}
as one of its main drawbacks.
Random phase shifts, without disentaglement of
the states, can also cause serious problems in an
ongoing quantum computation.
These are unitary transformations of the form:

\begin{equation}
\frac{1}{\sqrt {||a||^2 + ||b||^2}} \left[ \begin{array}{cc}
                                  a^* & b^* \\
                                  b & -a \\
                               \end{array} \right] 
\end{equation}
where $a^*$ and $b^*$ contain unknown phase angles.

Recent quantum computation algorithms\cite{Sh97,Gr97} use a method
of increasing the amplitude of a marked solution state
at the expense of unmarked states.
This is achieved by changing the
difference in the phase angles of
the marked and the unmarked states. 
But injection of random phases makes it
impossible to perform search that will exploit
quantum parallelism.

At the implementation level,
the representation of a unitary transformation
in terms of a sequence of small-degree gates (such as 2-bit gates)
will introduce random phase shifts at each gate that will
have an effect similar to random phases in a register.
In other words, the current conception of quantum computers
appears to be unsuitable in terms of implementation.

\section{State preparation}
A quantum register of length $n$ is postulated
where all the superpositions of the $N=2^n$ basis states exist with
the amplitudes:

\begin{equation}
(\frac{1}{\sqrt N},\frac{1}{\sqrt N},\frac{1}{\sqrt N} . . .\frac{1}{\sqrt N})
\end{equation}

The idea in the initialization of the quantum register is
to place all the $N=2^n$ states in a superposition
where each basis state is equally probable.
But how is this done?
By placing a bit, say a $0$, in each cell.
Now, the following transformation

\begin{equation}
M = \frac{1}{\sqrt 2} \left[ \begin{array}{cc}
                                  1 & 1 \\
                                  1 & -1 \\
                               \end{array} \right] 
\end{equation}
is applied to each bit, transforming it into the
superposition with the amplitudes $(
\frac{1}{\sqrt 2}, \frac{1}{\sqrt 2})$.

But this would be true only to within an unknown phase angle.
Strictly speaking, the state of the cell should be written as:

\begin{equation}
 |\phi\rangle = \frac{e^{i \theta }}{\sqrt 2}( |0\rangle +  |1\rangle) 
\end{equation}
where $\theta$ is the unknown phase of the initial
$0$.

Such a preparation for each qubit leads to the lumping together
of the uncertainty for the state of the register.
Here we can imagine that photons have been passed through a
horizontal polarizing filter and then rotated by
the transformation $M$ to produce superposition qubits.

But to consider this problem from a less idealistic
perspective, it should be remembered that the object
that carries the qubit, be it a photon, an electron, or an atom
or a molecule with a certain spin state, is already
physically present at its location.
Given that fact, the initialization procedure is to
let the object relax to the superposition state which in its
most general form will be:

\begin{equation}
 |\phi\rangle = \frac{e^{i \theta_{11} }}{\sqrt 2} |0\rangle +
 \frac{e^{i \theta_{12} }}{\sqrt 2} |1\rangle 
\end{equation}

The state of the quantum register will then be:

\begin{equation}
 |\phi\rangle = \frac{e^{i \theta_{1} }}{\sqrt N} |0...0\rangle +
 \frac{e^{i \theta_{2} }}{\sqrt N} |0...1\rangle + . . . +
 \frac{e^{i \theta_{N} }}{\sqrt N} |1...1\rangle 
\end{equation}
 
Although each of the $2^n$ states  has the same probability, 
the associated phases are unknown and so it is impossible
to use a method of amplitude amplification on any
marked state.
Phase rotation for a case where the phases are randomly
distributed will be meaningless.

\section{Quantum gates}
It is normal to speak of the phase function with the
state of the quantum system, but this can also be
expressed, equivalently, in terms of an arbitrary
phase associated with the unitary operator
because, operationally, from the point of view
of a measurement, these two are indistinguishable.
Clearly, the problem of the initialization of
the quantum register will have a parallel in the
initialization of the apparatus used to
implement unitary transformations.

The quantum computers 
implements the time-evolution operator $U$, that represents
the transformation on the data in the register, in terms
of smaller gates.
For example, DiVincenzo\cite{Di95} showed that two-bit
gates are universal for quantum computation.
This was done by showing that appropriate sequence of
two-bit gates can realize Deutsch's three-gates that
implement the Toffoli reversible gates.

But the unitary
transformation with each gate, in itself, is associated with
an unknown phase, and these values will migrate in the
direct product operation used to construct the larger 
gates.
In other words, the realization of the system
unitary matrix in terms of the small gates will be correct
only in the absence of the random phases.
In the recursive development of the S-matrices for the
various gates, Deutsch\cite{De89} failed to include 
the unknown phase with the embedded S-terms, assuming
thereby that the gates were initialized.

\section{Random phase shifts and decoherence}
Implementations 
of quantum computing  based on
trapped ions, quantum dots, cavity-QED, and
nuclear magnetic resonance (NMR) are 
being investigated\cite{Ek96}.
Here it is assumed that there are no randomized
phases in the initialized register.
But we must consider the
issue of decoherence that leads
to a decay of the superposition of states to a
particular base state due to an interaction with
the environment.
The decoherence time can vary from $10^{-12}$ sec for
electron-hole excitation in the bulk of a
semiconductor to $10^4$ sec for nuclear spin
on a paramagnetic atom.
If $t_d$ is the typical decoherence time, the
decoherence characteristics for a single qubit
are proportional to $e^{-t/t_d}$.

For multiple qubits one must multiply the
individual characteristics.
For a quantum register of $n$ qubits, the
decoherence characteristics are given by

\begin{equation}
e^{-tn/t_d} .
\end{equation}

In other words, the effective decoherence time
decreases linearly with the length of the register.

Decoherence may be viewed as a decay in the off-diagonal
elements of the density matrix representation of the
state of the register.
But in operational terms, the process is a cumulative
effect of random phase shifts introduced by the
interaction with the atoms of the environment.
Ultimately, the qubit object falls in one of
the basis states in equilibrium with the
state function of the environment.

In this perspective, the equilibrium may
be viewed to be the end result of a
walk executed by the phase of each
qubit within the energy state basin
to its least value.
But since the qubits are physically isolated to the
extent possible, one may take this walk to be a random one.
If the step size in this walk is $s$,
associated with a characteristic time of $\tau$,
then after $m\tau$ will be $m s^2$.

Since, the distribution of the random walk can be
approximated by the Gaussian function, a Gaussian
error with a linearly increasing variance will
characterize the departure from the desired values
of the phase angles.

To stress why the knowledge of relative phases is
important consider 
Grover's quantum search algorithm\cite{Gr97}, where
a certain transformation is applied to the state which
computes the property that the database item being searched
uniquely satisfies, marking that state in the process, further generating
transformed 
states in superposition.
Next, is the procedure that increases the amplitude of
the marked state progressively: the phase angle of the marked state is rotated through
$\pi$ radians and the diffusion transform $D$ applied as follows:

\begin{equation}
D_{ij} = \frac{2}{N} ~~ if  ~~i \neq j ~~ \& 
D_{ii} = -1 + \frac{2}{N}.
\end{equation}

This process is repeated a total of about $\frac{\pi}{4}\sqrt N$ times
after which the state is measured when it is found in
the marked state with probability close to 1, thus allowing us
to find the database item in about $\sqrt N$ steps compared
to the average of $\frac{N}{2}$ steps in a classical
algorithm.

In this algorithm, an error of $\epsilon$ in the phase of
the marked state will cause a corresponding error
of $\frac{1}{\sqrt N}(-\epsilon + \frac{2\epsilon}{N})$ in the amplitude
of the marked state, if it is assumed that the errors
in the phases of the other states cancel out.

When $N$ is large, the error in amplitude will be
$-\epsilon/\sqrt N$
in each step, but this will progressively increase
in subsequent steps.

\section{Quantum error correcting codes}

In a classical information system the basic error is represented by a
$0$ becoming a $1$ or vice versa.
The characterization of such errors is in terms of
an error rate, $\epsilon$,
associated with such flips.
The correction of such errors is achieved by appending 
check bits to a block of information bits.
The redundancy provided by the check bits can be
exploited to determine the location of errors using the
method of syndrome decoding.
These codes are characterized by a certain capacity
for error-correction per block.
Errors at a rate less than the capacity of the
code are {\it completely} corrected.

Now let us look at a quantum system.
Consider a single cell in a quantum register.
The error here can be due to a random unitary
transformation or by entanglement with the environment.
These errors cannot be defined in a graded sense because
of the group property of unitary matrices
and the many different ways the entanglements
can be expressed.
Let us consider just the first type of error,
namely that of random unitary transformation.
If the qubit is the state $| 0\rangle$, it can
become 
$a |0\rangle + b | 1 \rangle$.
Likewise, the state $| 00\rangle$ can become
$a | 00\rangle + b | 01 \rangle + c | 10 \rangle + d | 11\rangle $.
In the initialization of the qubit a similar
error can occur\cite{Ka98b}.
If the initialization process consists of collapsing
a random qubit to the basis state $| 0\rangle$, the definition
of the
basis direction can itself have a small error associated
with it. 
This error is analog and so, unlike error in classical
digital systems, it cannot be controlled.
In almost all cases, therefore, the qubits will have
entangled states, although the level of entanglement
may be very low.

From another perspective, classical error-correction codes
map the information bits into codewords in a higher
dimensional space so that if just a few errors occur in the
codeword, their location can, upon decoding, be identified.
This identification is possible because the errors perturb
the codewords, {\it locally}, within small spheres.
Quantum errors, on the other hand, perturb the information
bits, in a {\it nonlocal} sense, to a superposition of many states so the concept of
controlling all errors by using a higher dimensional
codeword space cannot be directly applied.

According to the positivist understanding of
quantum mechanics, it is essential to speak
from the point of view of the observer
and not ask about any intrinsic information in
a quantum state\cite{Ka98a}.
Let's consider, therefore, the representation of
errors by means of particles in a
register of $N$ states.

We could consider errors to be
equivalent to either $n$ bosons or fermions.
Bosons, in a superpositional state,
follow the Bose-Einstein statistics.
The probability of each pattern will
be given by

\begin{equation}
\frac{1}{\left( \begin{array}{c}
                N + n - 1 \\ n
                \end{array}
                  \right) }.
\end{equation}
 
So if there are 3 states and 1 error particle, we can only distinguish
between 3 states: $00,~01~or~10,~11$.
Each of these will have a probability of $\frac{1}{3}$.
To the extent this distribution departs from
that of classical mechanics, it represents nonlocality at
work.

If the particles are fermions, then they are
indistinguishable, 
and with $n$ error objects in $N$ cells, we have
each with the probability

\begin{equation}
\frac{1}{\left( \begin{array}{c}
                N  \\ n
                \end{array}
                  \right) }.
\end{equation}

If states and particles have been identified,
these statistics will be manifested by a group
of particles.
If the cells are isolated then their histories
cannot be described by a single unitary transformation.

Like the particles, the errors will also be
subject to the same statistics.
These statistics imply that the errors will not
be independent, an assumption that is basic to
the error-correction schemes examined in the
literature.

To summarize, important characteristics of quantum errors
that must be considered
are {\it component proliferation},
{\it nonlocal effects} and {\it amplitude error}.
All of these have no parallel in the classical case.
Furthermore,
quantum errors are analog and so the system cannot
be shielded below a given error rate. Such shielding
is possible for 
classical digital systems.

We know that a computation need not require
any expenditure of energy if it is cast in the form
of a reversible process.
A computation which is not reversible must involve
energy dissipation.
Considering conservation of
information+energy to be a fundamental principle, a
correction of random errors in the qubits
by unitary transformations, without any expenditure
of energy, violates this principle.

Can we devise error-correction coding for
quantum systems? To examine this,
consider the problem of protein-folding, believed to 
be NP-complete,
which is, nevertheless, solved efficiently by
Nature.
If a quantum process is at the basis of this
amazing result, then it is almost certain that
reliable or fault-tolerant quantum computing must exist
but, paying heed to the above-mentioned conservation law,
it appears such computing will require some lossy operations.

\section{Representing quantum errors}

Every unitary matrix can be transformed by a suitable
unitary matrix into a diagonal matrix with all its
elements of unit modulus.
The reverse also being true, quantum errors can play havoc.

The general unitary transformation representing errors
for a qubit is:

\begin{equation}
\frac{1}{\sqrt {||e_1||^2 + ||e_2||^2}} \left[ \begin{array}{cc}
                                  e_1^* & e_2^* \\
                                  e_2 & -e_1 \\
                               \end{array} \right] .
\end{equation}

These errors ultimately change the probabilities of
the qubit being decoded as a $0$ and as a $1$.
From the point of view of the user, when the quantum state
has collapsed to one of its basis states,
it is correct to speak of an error rate.
But such an error rate cannot be directly applied to
the quantum state itself.

Unlike the classical digital case, quantum errors cannot
be completely eliminated because they are essentially analog
in nature.

The unitary matrix (11) represents
an infinite number of cases of error.
The error process is an analog process, 
and so, in general, such errors cannot be corrected.
From the point of view of the qubits, it is a
nonlocal process.

If it is assumed that the error process can be represented
by a small rotation and the initial
state is either a $0$ or a $1$, then this rotation
will generate a superposition of the two states 
but the relative amplitudes will be different
and these could be exploited in some specific situations
to determine the starting state.
But, obviously, such examples represent trivial cases.

The error process may be usefully represented 
by a process of quantum diffusions and phase
rotations.

Shor\cite{Sh95} showed how the decoherence
in a qubit could be corrected by a system
of triple redundancy coding where each qubit is
encoded into nine qubits as follows:

\[|0\rangle \rightarrow \frac{1}{2\sqrt 2}  
( |000\rangle + | 111\rangle )
( |000\rangle + | 111\rangle )
( |000\rangle + | 111\rangle )\],

\begin{equation}
|1\rangle \rightarrow \frac{1}{2\sqrt 2}  
( |000\rangle - | 111\rangle )
( |000\rangle - | 111\rangle )
( |000\rangle - | 111\rangle ).
\end{equation}

Shor considers the decoherence process to be one where
a qubit decays into a weighted amplitude
superposition of its basis states.
In parallel to the assumption of independence of
noise in classical information theory,
Shor
assumes that only one qubit out of the
total of nine decoheres.
Using
a Bell basis, Shor then shows that one can determine the
error and correct it.

But this system does not work if more than one
qubit is in error.
Quantum error is analog, so each qubit will be
in some error and so this scheme will, in practice,
not be useful in {\it completely} eliminating
errors.

The question of decoherence, or error, must be considered as
a function of time. 
One may use the exponential function $\lambda e^{-\lambda t}$
as a measure of the decoherence probability of the
amplitude of the qubit.
The measure of decoherence that has taken place by time $t$
will then be given by the probability, $p_t$:

\begin{equation}
p_t = 1 - \lambda e^{-\lambda t}.
\end{equation}

In other words, by time $t$, the amplitude of the
qubit would have decayed to a fraction
$(1 - \lambda e^{-\lambda t})$ of its original value.
At any time $t$, there is a $100 \%$ chance that the
probability amplitude of the initial state will
be a fraction $\alpha_k < 1$ of the
initial amplitude.

If we consider a rotation error in each qubit through angle $\theta$,
there exists some $\theta_k$ so that the probability

\begin{equation}
Prob ( \theta > \theta_k) \rightarrow 1.
\end{equation}

This means that we cannot represent the qubit error
probability by an assumed value $p$ as was done
by Shor in analogy with the classical case.
In other words, there can be no guarantee of
eliminating decoherence.

\section{Recently proposed error-correction codes}

The recently proposed models of
quantum error-correction codes assume
that the error in the 
qubit state $a |0\rangle + b | 1 \rangle$
can be either a bit flip $ |0 \rangle \leftrightarrow | 1 \rangle$,
a phase flip between the relative phases of $| 0\rangle$
and $| 1 \rangle $, or both \cite{St96,Sh96,Pr97}.

In other words, the errors are supposed to take the pair 
of amplitudes $(a,b)$
to either $(b,a)$, $(a, -b)$, or $(-b,a)$.

But these three cases represent a vanishingly small subset
of all the random unitary transformations associated
with arbitrary error.
These are just three of the infinity of rotations 
and diffusions that the
qubit can be subject to.
The assumed errors,
which are all local,
do not, therefore, constitute a distinguished set on
any physical basis.

In one proposed error-correction code,
each of
the states
$ |0\rangle $ or $  | 1 \rangle$
is represented by
a 7-qubit code, where the 
strings of the codewords represent the
codewords of the single-error correcting
Hamming code, the details of which we don't
need to get into here.
The code for $| 0 \rangle$ has an even number of
$1$s and the code for $|1 \rangle$ has an odd number
of $1$s.

\[| 0\rangle_{code} = \frac{1}{\sqrt8} (|0000000\rangle + |0001111\rangle+
|0110011\rangle + |0111100\rangle\\\]

\begin{equation}
 + |1010101\rangle
+|1011010\rangle +| 1100110\rangle + |1101001\rangle),
\end{equation}

\[|1\rangle_{code} = \frac{1}{\sqrt8} (|1111111\rangle + |1110000\rangle+
|1001100\rangle + |1000011\rangle \\\]

\begin{equation}
+ |0101010\rangle
+|0100101\rangle +| 0011001\rangle + |0010110\rangle).
\end{equation}

As mentioned before, the errors are assumed to be either in terms of
phase-flips or bit-flips.
Now further ancilla bits--- three in total--- are augmented that compute the
syndrome values.
The bit-flips, so long as limited to one in each group, can
be computed directly from the syndrome.
The phase-flips are likewise computed, but only after a change of
the bases has been performed.

Without going into the details of these steps, which are
a straightforward generalization of classical error
correction theory, it is clear that the assumption
of single phase and bit-flips is completely artificial.

In reality, errors in the 7-qubit words will generate an
superposition state of 128 sequences,
rather than the 16 sequences of equations (15) and (16), together with
16 other sequences of one-bit errors, where the errors
in the amplitudes are limited to the phase-flips mentioned
above.
{\it All kinds of bit-flips}, as well as
modifications of the amplitudes will be a part of
the quantum state.

We can represent the state, with the appropriate
phase shifts associated with
each of the 128 component states, as follows:

\begin{equation}
 |\phi\rangle = e^{i \theta_{1} } a_1 |0000000\rangle +
 e^{i \theta_{2} } a_2 |0000001\rangle + . . . +
 e^{i \theta_{N} } a_N |1111111\rangle)
\end{equation}

While the amplitudes of the newly generated components
will be small, they would, nevertheless, have a
non-zero error probability.
These components, cannot be corrected by the code
and will, therefore, contribute to an residual
error probability.

The amplitudes implied by (17) will, for the 16 sequences
of the original codeword
after the error has enlarged the set, be somewhat different from 
the original values.
So if we speak just of the 16 sequences
the amplitudes cannot be preserved without error.

Furthermore, the phase errors in (17) cannot be corrected.
These phases are of crucial importance in 
many recent quantum algorithms.

It is normally understood that in classical systems if
error rate is smaller than a certain value, the error-correction
system will correct it.
In the quantum error-correction systems, this important
criterion is violated.
Only certain specific errors are corrected, others even
if smaller, are not.

In summary,
 the proposed models are based on a
local error model while real errors
are nonlocal where we must consider the issues of
component proliferation and amplitude errors.
These codes are not capable of completely
correcting small errors that cause
new component states to be created.

\section{The sensitivity to errors}
The nonlocal nature of the quantum errors is seen
clearly in the sensitivity characteristics of these
errors.

Consider that some data sets related to a problem are being
simultaneously processed by
a quantum machine.
Assume that by some process of phase switching and diffusion
the amplitude of the desired solution out of the entire set is slowly
increased at the expense of the others.
Nearing the end of the computation, the sensitivity of the
computations to errors will increase dramatically,
because the errors will, proportionately, increase for
the smaller amplitudes.
To see it differently, it will be much harder to reverse the
computation if the change in the amplitude or phase
is proportionally greater.

This means that the
``cost'' of quantum error-correction will depend on the
state of the computing system.
Even in the absence of errors, the sensitivity
will change as the state evolves,
a result, no doubt, due to the nonlocal nature of quantum errors.
These errors can be considered to be present 
at the stage of state preparation and through
the continuing interaction with the environment
and also due to the errors in the applied
transformations to the data.
In addition, there may exist nonlocal correlations
of qubits with those in the environment. The
effect of such correlations will be unpredictable.

Quantum errors
cannot be localized. For example,
when speaking of rotation errors, there always exists some
$\theta_k > 0$  so that $prob (\theta > \theta_k) \rightarrow 1$.

When doing numerical calculations on a computer, it is
essential to have an operating regime that provides
reliable, fault-tolerant processing.
Such regimes exist in classical computing.
But the models currently under examiniation for 
quantum computing cannot eliminate
errors completely.
The method of
syndrome decoding, adapted from the
theory of classical error-correcting codes,
appears not to be the answer to the problem of fault-tolerant
quantum computing.
New approaches to error-correction may be needed.

\section{Conclusions}
The undetermined phase of a quantum
state can be seen, equivalently, in an undetermined
phase associated with each unitary operator.
Normally, this has no significance because the
usual representations deal with the entire system and so
the phase is effectively a lumped term that has no
observational value.
In considering a unitary transformation 
as being built out of smaller blocks, 
the phase cannot be ignored.
In other words, there is no simple way we can effectively
``initialize'' each quantum gate.

If one did not concern oneself with the question of
the realization of the gates, assuming that the
system unitary transformation will be somehow
carried out, one still has a difficulty with
the random phases in the component states of a
quantum register.
Given these random phases, one cannot manipulate
the amplitudes to increase the value
for a marked state as is
required in the search problem.
If the random phases exist in the initialized register,
computations exploiting quantum superposition cannot
be performed.
If the randomization doesn't exist in the
initialized register and is forced upon the
computation in the later stages, then this might
shorten the time range where useful computations
can be performed even more than by decoherence.

We don't know of any
simple method to correct for the random
phase errors by the use of 
the proposed quantum error correction
codes because these errors will not necessarily be
within bounds\cite{Pr97}.
Besides, the error-correction systems will be plagued
with the same random phase problems that apply to
other quantum gates and registers.
As mentioned before, it is
nonlocality, related both to the evolution of the
quantum information system and errors, 
that makes it difficult for
syndrome decoding to work.

How should error-correction be defined then?
Perhaps through a system akin to associative
learning in spin glasses.

\end{document}